\documentclass[aps,prl,10pt,twocolumn,preprintnumbers,superscriptaddress,showpacs,floatfix,nofootinbib]{revtex4-2}
\usepackage{textcomp,gensymb}
\usepackage{hyperref}
\usepackage{graphicx}
\usepackage{amsfonts,amsmath,amssymb,bm,bbm}
\usepackage{color,soul}
\usepackage{xcolor}
\usepackage{slashed}
\usepackage[toc,page]{appendix}
\usepackage{flushend}
\usepackage{physics}
\usepackage{graphicx}
\usepackage{dcolumn}
\usepackage{float}

\hypersetup{
    pdfnewwindow=true,      
    colorlinks=true,       
    linkcolor=blue,          
    citecolor=blue,        
    filecolor=blue,      
    urlcolor=blue        
}

\begin{document}

\title{A New Probe of Strongly-Interacting Dark Sector using Neutrino Telescopes}

\author{Jose A. Macias Cruz}
\affiliation{Department of Physics, Washington University,
St. Louis, MO 63130, USA}

\author{Christopher V. Cappiello}
\affiliation{Department of Physics, Washington University,
St. Louis, MO 63130, USA}
\affiliation{McDonnell Center for the Space Sciences, Washington University,
St. Louis, MO 63130, USA}
\author{P. S. Bhupal Dev}
\affiliation{Department of Physics, Washington University,
St. Louis, MO 63130, USA}
\affiliation{McDonnell Center for the Space Sciences, Washington University,
St. Louis, MO 63130, USA}
\author{Suchita Kulkarni}
\affiliation{Institute of Physics, NAWI Graz, University of Graz, Universit\"atsplatz 5, A-8010 Graz, Austria}

\begin{abstract}
A confining dark sector modeled after quantum chromodynamics provides a well-motivated framework for strongly-interacting dark matter. We show that dark vector mesons in a confining dark sector can be resonantly produced in high-energy neutrino scattering on the cosmic neutrino background, generating a distinctive absorption feature in the cosmic neutrino energy spectrum. Unlike conventional attenuation effects, this feature directly reflects the mass spectrum of dark resonances and can therefore serve as a spectroscopic signature of a strongly-interacting dark sector. We find that a broad range of dark-sector parameter space is within the reach of future neutrino telescopes, including IceCube-Gen2 Radio. Our proposal thus turns the cosmic neutrino background into a target for discovering dark-sector resonances, while simultaneously providing a novel probe of dark-sector interactions with neutrinos. 
\end{abstract}

\maketitle

\section{Introduction}
Identifying the nature and properties of dark matter (DM) is among the most pressing questions in particle physics and cosmology. Among the many DM models proposed~\cite{Bertone:2018krk}, a particularly well-motivated framework, closely paralleling the origin and stability of ordinary matter in the Standard Model (SM), is a strongly-interacting dark sector (DS) with its own confining dynamics and stabilizing symmetries. The confining dark sector  produces a tower of bound states, and the lightest one (the dark pion $\pi_D$) may serve as a DM candidate~\cite{Hochberg:2014kqa}. These scenarios naturally realize a light (sub-GeV-scale) strongly-interacting massive particle (SIMP) DM whose relic density is generated by number-changing processes within the DS~\cite{Hochberg:2014kqa, Hochberg:2018vdo, Choi:2018iit, Heikinheimo:2018esa, Bernreuther:2019pfb, Katz:2020ywn, Bernreuther:2023kcg, Braat:2023fhn, Pomper:2024otb, Chu:2024rrv, Davighi:2024zip, Belanger:2026ctm}, rather than their interactions with the SM particles. The SIMP DM is therefore difficult to observe in the traditional DM searches, which mostly rely on DM coupling weakly with the SM. For a review of the SIMPs, their detection strategies and present status,  see Refs.~\cite{Cline:2021itd, Albouy:2022cin, Asadi:2026mip}. This situation presents a unique challenge, where on one hand a tower of DS particles may be within reach because they are light, but on the other hand, their feeble interactions with the SM make them hard to detect.

In this work, we present a new way to detect such light, strongly-interacting DS particles, based on two observations. The first is that the dark vector mesons (the dark rho $\rho_D$ being the lightest) can couple to the SM neutrinos via their mixing with the SM $Z$-boson or a new  $Z^\prime$ gauge boson~\cite{Lee:2015gsa,Hochberg:2017khi, Berlin:2018tvf, Bernreuther:2019pfb, Bernreuther:2025xqk}. The second is that the Universe is filled with the cosmic neutrino background (C$\nu$B), which has the highest flux of any known neutrino population and provides a natural target for DS-neutrino interactions. It is therefore conceivable that high-energy neutrinos may resonantly annihilate with the C$\nu$B via the dark rho meson, which then promptly decays to dark pions, thus producing a distinct absorption feature in the high-energy neutrino spectrum, which would be detectable at current/future neutrino telescopes, such as IceCube/IceCube-Gen2~\cite{IceCube-Gen2:2020qha}.

Resonant absorption features for high-energy neutrinos have been widely explored, both in the context of the SM $Z$ boson~\cite{Weiler:1982qy, Roulet:1992pz,Yoshida:1996ie,Fargion:1997ft,Weiler:1997sh,Yoshida:1998it, Eberle:2004ua, Barenboim:2004di, Maitra:2025opp} and $\rho$ meson~\cite{Bander:1994tc,Paschos:2002sj,Dev:2021tlo,Brdar:2022kpu} resonances, and also in the context of non-standard neutrino interactions with light mediators~\cite{Ioka:2014kca,Ng:2014pca,Ibe:2014pja,Araki:2014ona,Cherry:2014xra,Kamada:2015era,DiFranzo:2015qea,Altmannshofer:2016brv,Bustamante:2020mep,Creque-Sarbinowski:2020qhz,Carpio:2021jhu,Esteban:2021tub,Carpio:2022lqk,delaVega:2024pbk}. 
Given existing constraints on the sum of neutrino masses~\cite{Planck:2018vyg,Elbers:2025vlz}, the neutrino energy required to produce the $Z$ resonance would be well above the GZK cutoff~\cite{Greisen:1966jv,Zatsepin:1966jv}. As for the vector meson resonances~\cite{Dev:2021tlo}, the lightest one, i.e. the $\rho$ meson, is the most relevant due to its large width-to-mass ratio, and its mass being a factor of $\sim$100 lighter than the $Z$ boson. This reduces the required neutrino energy by around 4 orders of magnitude, putting the resulting absorption feature in the energy range where several proposed experiments will have their best sensitivity to cosmic neutrinos~\cite{Ackermann:2022rqc}. However, given the extremely small branching ratio (BR) of rho into neutrinos (summed over all flavors)~\cite{Gao:2018seg}, 
\begin{align}
   {\rm BR}(\rho_{\rm SM}\to \nu\bar{\nu})= (2.41\pm 0.02)\times 10^{-13} \, ,
   \label{eq:BR}
\end{align}
the detection of a $\rho$-induced absorption feature requires the existence of a large C$\nu$B overdensity~\cite{Brdar:2022kpu}.

Our proposed signal has two main distinct features: (i) The resonance energy depends on the dark meson spectrum which, depending on the confining DS dynamics, can be different from the SM spectrum, and (ii) the strength of the absorption signal depends on ${\rm BR}(\rho_D\to \nu\bar{\nu})$, which can be much larger than the SM value~\eqref{eq:BR}, depending on the DS coupling to neutrinos. This allows us to produce a detectable absorption feature in the cosmic neutrino spectrum without requiring a C$\nu$B overdensity which can only enhance the signal. 
We show that future cosmic neutrino experiments will be sensitive to a wide range of the DS particle masses and couplings to neutrinos, while being compatible with the SIMP DM relic density requirement. 
Astrophysical uncertainties in the high-energy neutrino flux, and the uncertainty in the lightest neutrino mass, do not change the qualitative features of our proposal.

\section{Dark Sector Model}\label{sec:model}

We consider the SIMP paradigm~\cite{Hochberg:2014kqa}, i.e.~a strongly-interacting DS with $N_c$ colors and $N_f$ flavors, where the dark pion emerges as the pseudo-Nambu-Goldstone boson at low energy as a result of non-perturbative interactions and serves as a DM candidate.  Along with dark pions, the DS also contains a whole tower of dark mesons in analogy with QCD. In particular, we consider a $SU(N_c) \times U(1)_D$ extension of the SM, where a massive gauge boson associated with the broken $U(1)_D$ symmetry serves as the SM-DS mediator~\cite{Hochberg:2015vrg, Lee:2015gsa,Hochberg:2017khi, Berlin:2018tvf, Bernreuther:2019pfb, Bernreuther:2025xqk}. This $U(1)_D$ coupling also breaks the dark flavor symmetry in the DS and allows one of the diagonal dark rho mesons to mix with the mediator. For simplicity, we fix $N_c = 3, N_f = 3 $. As proposed in Ref.~\cite{Bernreuther:2023kcg}, we use fits to nonperturbative functional-method results~\cite{Maris:2005tt} to compute the relative dark pion-rho meson mass spectrum: 
\begin{equation}
    \xi \equiv \frac{m_{\pi_D}}{f_{\pi_D}} = 7.79\dfrac{m_{\pi_D}}{m_{\rho_D}} + 0.57 \left(\dfrac{m_{\pi_D}}{m_{\rho_D}} \right)^2 \, ,
    \label{eq:xi}
\end{equation}
where $f_{\pi_D}$ is the dark pion decay constant, and $m_{\pi_D}, m_{\rho_D}$ are the dark pion and rho masses, respectively. Eq.~\eqref{eq:xi} gives a relation between $f_{\pi_D}, m_{\pi_D}$ and $m_{\rho_D}$, and is valid for $1<m_{\rho_D}/m_{\pi_D}\lesssim 20$ which corresponds to $8.4>\xi\gtrsim 0.4$. 
We restrict ourselves to $\xi \lesssim 4$ corresponding to $m_{\rho_D}>2m_{\pi_D}$ such that the $\rho_D \to \pi_D \pi_D$ decay mode is allowed with the decay rate  
\begin{align}
\Gamma_{\rho_D}=\frac{\abs{g_{\rho_D\pi_D\pi_D}}^2}{48\pi}m_{\rho_D}\left(1 - \frac{4m_{\pi_D}^2}{m_{\rho_D}^2}\right)^{3/2} . 
\end{align}
Here $g_{\rho_D\pi_D\pi_D}$ is the effective $\rho_D\pi_D\pi_D$ coupling, which we take as $\approx 6$ as in QCD (see e.g., Ref.~\cite{Xie:2008ts}).
While ${\rm BR}(\rho_D\to \pi_D\pi_D)$ is almost 100\% when kinematically allowed, the small $\rho_D - Z^\prime$ mixing allows for $\rho_D \to \nu \bar\nu$ decay mode which we exploit in our analysis. ${\rm BR}(\rho_D\to \nu\bar{\nu})$ is not uniquely determined by the choice of $m_{\pi_D}$ and $\xi$, and depends on the details of the model construction. We do not attempt to establish a full model but rather treat ${\rm BR}(\rho_D\to \nu\bar{\nu})$ as a free parameter to keep our analysis generic. Depending on the model construction, e.g. in neutrinophilic $Z'$ models~\cite{Farzan:2016wym,Nomura:2017wxf,Farzan:2017xzy,Bakhti:2018avv,Chauhan:2020mgv,Abdallah:2021npg,Chauhan:2022iuh}, ${\rm BR}(\rho_D\to \nu\bar{\nu})$ can be as large as $\sim 10^{-4}$, which we use as an upper limit in our analysis.

\section{Dark Rho Resonance}\label{sec:neutrinos}

We consider high-energy neutrino scattering off C$\nu$B to produce the DS particles. The cross section for a $\nu\bar{\nu}$ collision to produce a dark-rho resonance is given by the Breit-Wigner formula~\cite{ParticleDataGroup:2026mpi}:
\begin{equation}
    \sigma({\nu \overline{\nu} \rightarrow \rho_D}) = \frac{48\pi}{m_{\rho_D}^2} \frac{s\Gamma_{\rho_D}^2 \mathrm{BR}(\rho_D \rightarrow \nu \overline{\nu})}{(s - m_{\rho_D}^2)^2 +s^2\Gamma_{\rho_D}^2/m_{\rho_D}^2}\,.
    \label{eq:cross}
\end{equation}
Here $s=2m_{\nu}E$ is the squared center-of-momentum frame energy, $E$ is the incoming energy of the cosmogenic neutrino, and the C$\nu$B is assumed to be non-relativistic. Thus, the neutrino energy required to produce a dark rho resonance is $E_{{\rm res}} = m_{\rho_D}^2/2m_{\nu}$. We adopt a benchmark value of $m_{\nu, {\rm min}} = 0.01$ eV for the lightest neutrino mass unless otherwise specified, and assume normal ordering for the other two neutrino mass eigenstates.

\section{Cosmic Neutrino Spectrum}

In the absence of any non-standard neutrino interactions, the flux of ultra-high energy neutrinos at Earth is given by~\cite{Ahlers:2009rf}
\begin{equation}
    \Phi_0(E) = \frac{1}{4\pi}\int_0^{\infty}\textrm{d}z\frac{1}{H(z)}L(z,(1+z)E))\,,
    \label{eq:phi0}
\end{equation}
where $H(z)$ 
is the Hubble expansion parameter, $E'(z,E)$ is the energy at which a neutrino must be produced at redshift $z$ in order to have an energy $E$ at redshift $z=0$. $L(z,E)$ is the source spectrum of ultra-high energy neutrinos at redshift $z$ which can be parameterized as
\begin{equation}
    L(z,E) = F(z)L(0,E)\,,
\end{equation}
where $F(z)$ encapsulates the source redshift evolution. For concreteness, as in Refs.~\cite{Ahlers:2009rf, Muzio:2023skc}, we consider an  observationally-informed AGN evolution~\cite{Hasinger:2005sb,Stanev:2008un}:
\begin{align}
F_{\rm AGN}(z) = \begin{cases}
    (1+z)^5 & z < 1.7\\
    (1+1.7)^5 & 1.7 < z < 2.7\\
    (1+1.7)^5 e^{2.7-z} & z > 2.7\,.
\end{cases}
\label{eq:AGN}
\end{align}
We take the luminosity function to be a sum of two terms, each having the functional form of the test functions introduced in Ref.~\cite{Ahlers:2009rf}, one arising from neutron decay and another from muon decay:
\begin{multline}
    L(0,E) = L_\mu\left(\frac{E}{\epsilon E_\mu}\right)^{-1}\exp{-\left(\frac{E}{\epsilon E_\mu}\right)^{0.48}} 
    \\+ L_n\left(\frac{E}{\epsilon E_n}\right)^{-1}\exp{-\left(\frac{E}{\epsilon E_n}\right)^{3.5}}\,,
    \label{5}
\end{multline}  
where $\epsilon = 0.07$ is the ratio of the average neutrino energy to average neutron energy~\cite{Ahlers:2005sn}. The parameters $L_\mu = 6.1\times10^{-58}\,\rm{cm^{-3}\,s^{-1}\,sr^{-1}\,GeV^{-1}}$, $L_n = 4.3\times10^{-57}\,\rm{cm^{-3}\,s^{-1}\,sr^{-1}\,GeV^{-1}}$ and $E_\mu = 2.1\times10^{11}$ GeV, $E_n = 2.9\times10^{10}$ GeV have been determined by fitting the predicted spectrum from Ref.~\cite{Muzio:2023skc} (their Fig.~7 left panel, green-dashed curve) using an AGN-informed source evolution [cf.~Eq.~\eqref{eq:AGN}] and using SIBYLL2.3C hadronic interaction model (HIM)~\cite{Fedynitch:2018cbl}. We checked that while   using another source evolution such as SFR~\cite{Hopkins:2006bw,Yuksel:2008cu,Robertson:2015uda} or GRB~\cite{Kistler:2007ud,Yuksel:2008cu}, or using another HIM such as EPOS-LHC~\cite{Pierog:2013ria} does lead to some variations in the neutrino flux prediction, the qualitative features of the attenuated spectrum as shown in our Fig.~\ref{fig:spectrum} remain unchanged.     

\section{Neutrino Absorption}
Analogous to the treatment in Ref.~\cite{Ahlers:2009rf}, the attenuated flux of ultra-high-energy neutrinos reaching the Earth can be written as
\begin{multline}
    \Phi(E) = \frac{1}{4\pi}\int_0^{\infty}\textrm{d}z\frac{1}{H(z)}L(z,(1+z)E)\\
   \quad  \times\textrm{exp}\left[-\int_0^{{\rm min}(z,z_{\rm cloud})} \textrm{d}z'
    \frac{\Gamma_{\rm ann}(z',(1+z')E)}{(1+z')H(z')}\right]
    \,,
    \label{eq:attenuated}
\end{multline}
where $\Gamma_{\rm ann}(z,E) = n_\nu(z)\sigma(z, E)$ is the rate of annihilation of neutrinos with energy $E$ at redshift $z$, $\sigma$ is the $\nu\bar\nu$ annihilation cross section for a given mass eigenstate, $n_{\nu}(z) = n_{\nu,0}\eta(1+z)^3$ is the C$\nu$B number density, with $n_{\nu,0}\simeq 56~{\rm cm}^{-3}$ being the local density per neutrino flavor as predicted by standard  cosmology, and $\eta$ accounts for any overdensity that might be present. Gravitational clustering leads to a local overdensity of $\eta\sim {\cal O}(1)$~\cite{Ringwald:2004np,Mertsch:2019qjv, Holm:2024zpr}, while large overdensities may be possible in presence of non-standard neutrino interactions~\cite{Wise:2014ola, Smirnov:2022sfo}. We assume a flavor-universal coupling to neutrinos, so the number density per flavor is appropriate to use for all high-energy neutrinos. We will set $\eta=1$ unless otherwise specified in which case the size of the C$\nu$B cloud, $z_{\rm cloud}$, is the cosmic horizon. When $\eta>1$, $z_{\rm cloud}$ is the redshift corresponding to the maximum size of the local C$\nu$B cloud so that the Universe is not overclosed.

In the presence of our DS-neutrino interactions, we can have resonant $\rho_D$ production from $\nu\bar\nu$ annihilation, followed by $\rho_D\to \pi_D\pi_D$ decay. This leads to an attenuated neutrino flux that depends on the DS parameters ${\rm BR}(\rho_D\to \nu\bar\nu)$ [cf.~Eq.~\eqref{eq:cross}] and $m_{\rho_D}$, which in turn is determined by $m_{\pi_D}$ and $\xi$ [cf.~Eq.~\eqref{eq:xi}].

\begin{figure}
    \centering
    \includegraphics[width=\linewidth]{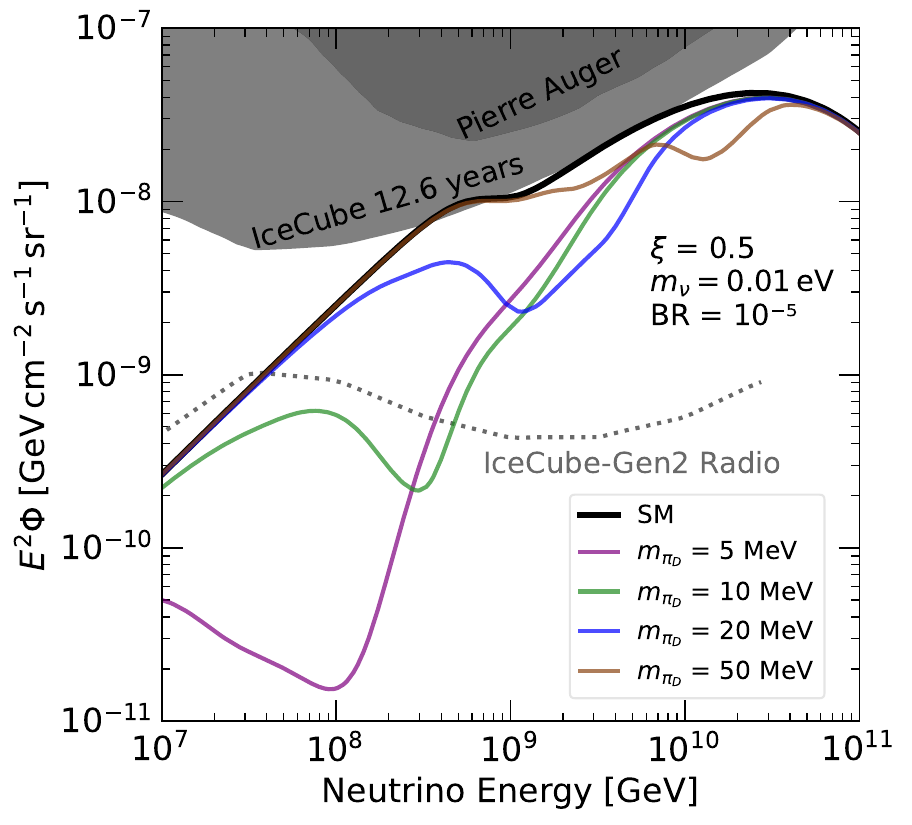}
    \caption{The cosmogenic neutrino flux without attenuation (solid, black) and with attenuation due to annihilation with the C$\nu$B (solid, colored), for several values of $m_{\pi_D}$. For comparison, we show the current constraints (gray shaded) from Pierre Auger~\cite{PierreAuger:2019ens} and IceCube~\cite{IceCubeCollaborationSS:2025jbi}, and future sensitivity of IceCube-Gen2 Radio~\cite{IceCubeGen2TDR}.}
    \label{fig:spectrum}
\end{figure}

To illustrate the effect of neutrino-DS interactions in this model, we compare the unattenuated and attenuated high-energy neutrino spectra given by Eqs.~\eqref{eq:phi0} and \eqref{eq:attenuated}  respectively. Figure~\ref{fig:spectrum} shows the effect of neutrino absorption by the C$\nu$B on the high-energy neutrino flux due to resonant $\rho_D$ production for different values of dark pion mass and for benchmark values of $\xi=0.5$ and ${\rm BR}(\rho_D\to \nu\bar\nu)=10^{-5}$. For the SM-expected neutrino flux, we see two distinct bumps, one arising from muon decay and the other from neutron decay [c.f. Eq.~\eqref{5}].  
Furthermore, we see that for a given $\xi$, as $m_{\pi_D}$ increases, the energy of the absorption feature increases, while decreasing $m_{\pi_D}$ shifts the feature to lower energies. This occurs because as $m_{\pi_D}$ increases or decreases, so does $m_{\rho_D}$, requiring a correspondingly higher or lower neutrino energy to undergo resonant annihilation. The depth of the dips is governed by ${\rm BR}(\rho_D\to \nu\bar{\nu})$ and $m_{\pi_D}$ in Eq.~\eqref{eq:cross}, while the width is governed by the $\rho_D$ width that is primarily determined by $\rho_D\to \pi_D\pi_D$. We see multiple resonant dips along the spectrum with the spread of dips on a single line being due to the three neutrino mass eigenstates. We show the existing flux limits (gray-shaded regions) from Pierre Auger~\cite{PierreAuger:2019ens} and   IceCube~\cite{IceCubeCollaborationSS:2025jbi}, and the projected sensitivity of IceCube-Gen2 Radio array~\cite{IceCubeGen2TDR}. We find that for the choice of parameters in Fig.~\ref{fig:spectrum}, the attenuated spectrum is well within the projected sensitivity.

\section{IceCube-Gen2 Radio Sensitivity}
We compute the detectability of the absorption feature in question using the proposed IceCube-Gen2 Radio array~\cite{IceCubeGen2TDR}, which would provide one of the most sensitive probes of the highest-energy cosmic neutrino spectrum in the considered energy range~\cite{Ackermann:2022rqc}. As part of IceCube-Gen2~\cite{IceCube-Gen2:2020qha}, a surface array of radio antennas is planned which will cover approximately 500 km$^2$ of Antarctic ice, with a total effective volume reaching $\sim$1600 km$^3$\,sr at an energy of 1 EeV~\cite{IceCube-Gen2:2021rkf, IceCube-Gen2:2023vtj,IceCubeGen2TDR}. The energy resolution of the radio array is expected to be approximately 65\% for energies above 100 PeV~\cite{IceCubeGen2TDR}. We assume a 10-year observation time, matching the sensitivity projection shown in Refs.~\cite{IceCube-Gen2:2020qha, IceCube-Gen2:2021rkf, IceCubeGen2TDR}. The flux sensitivity is shown in Fig.~\ref{fig:spectrum} (dotted curve), taken from the TDR~\cite{IceCubeGen2TDR}.  

For given values of the neutrino overdensity $\eta$, the $\rho_D$ BR to neutrinos, the dark pion mass $m_{\pi_D}$, and $\xi$, we compute the neutrino spectrum after absorption as shown in Fig.~\ref{fig:spectrum}. We find our attenuated and unattenuated number of events and scan across the entire energy range using overlapping bins of size $E_{\textrm{att}}/2$ to $2E_{\textrm{att}}$ to account for the energy resolution of the detector. We identify the energy bin within the entire energy range of interest,  $[10^7,10^{11}]$ GeV, where the attenuation signal is the most statistically significant. We then determine the degree to which this choice of model parameters would be ruled out at $2\sigma$, assuming that the number of observed events equals the number of expected events with no absorption, using a Feldman-Cousins~\cite{Feldman:1997qc} confidence interval approach.

\section{Results and Discussion}\label{sec:results}
\begin{figure}[t!]
    \centering
    \includegraphics[width=\linewidth]{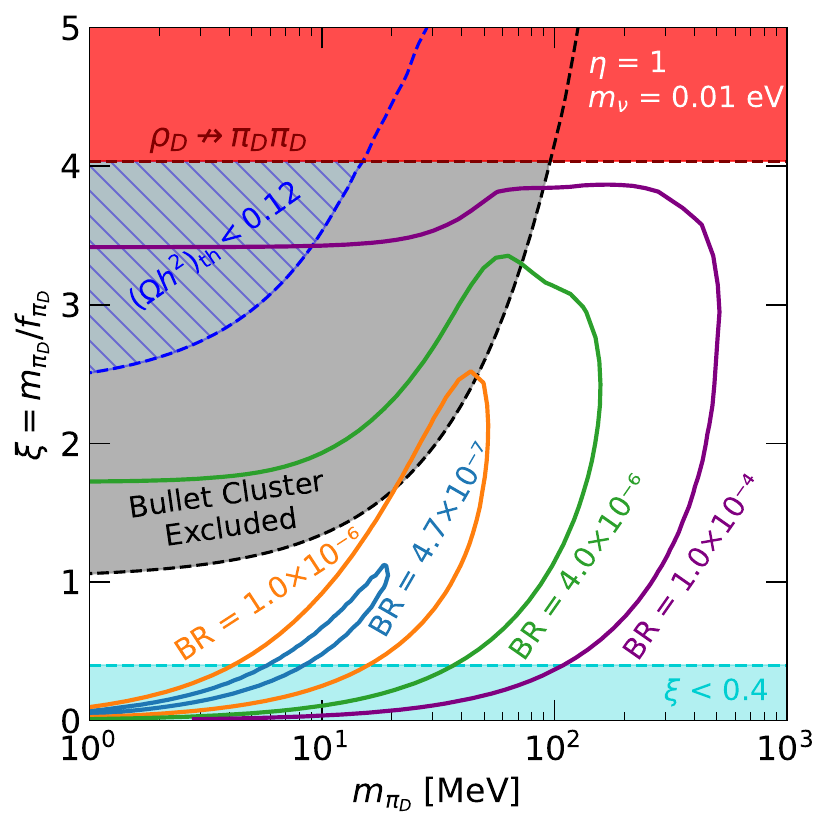}
    \caption{$2\sigma$ sensitivity contours of the absorption signal at IceCube-Gen2 Radio in the $m_{\pi_D}$--$\xi$   parameter space for different values of ${\rm BR}(\rho_D\to \nu\bar\nu)$. We also show the regions excluded by the thermal relic requirement (blue-hatched) and the Bullet Cluster limit on self-interacting DM (gray-shaded). The $\rho_D\to \pi_D \pi_D$ kinematic cutoff is shown by the red-shaded region, while the cyan-shaded region is outside the validity range of Eq.~\eqref{eq:xi}.}
    \label{fig:BR_contours-eta=1}
\end{figure}

Fig.~\ref{fig:BR_contours-eta=1} shows the regions of the $m_{\pi_D}$--$\xi$  parameter space to which IceCube-Gen2 Radio will be sensitive at $2\sigma$, for several choices of ${\rm BR}(\rho_D\to \nu\bar{\nu})$. With the standard C$\nu$B number density,~i.e. with $\eta=1$, we find that the minimum BR that can be probed at $2\sigma$ is around $4.7\times 10^{-7}$. In presence of a large C$\nu$B overdensity $\eta\gg 1$, much smaller values of BR can be probed; see End Matter  Fig.~\ref{fig:BR_contours-eta=1e11 and Eta_contours}. 
As we increase the BR, the absorption signal strength increases and more parameter space can be probed.  
We also show existing limits (shaded regions)~\cite{Braat:2023fhn} from the thermal relic requirement (blue-hatched) and from Bullet Cluster data which constrains  self-interacting DM (gray-shaded). In the red-shaded region with $\xi\gtrsim 4$, the 2-body decay $\rho_D \rightarrow \pi_D \pi_D$ is kinematically forbidden, while the cyan-shaded region lies outside the range of validity of Eq.~\eqref{eq:xi}.

As shown in Fig.~\ref{fig:BR_contours-eta=1}, the DM relic density obtained by assuming a SM--DS kinetic equilibrium is disfavored by the Bullet Cluster constraints. Nevertheless, the analysis presented here provides a useful probe of strongly-interacting DS. First, if the SM--DS kinetic equilibrium is not obtained: this can happen when the coupling between the two sectors is very feeble; such a scenario has been investigated e.g. in Ref.~\cite{Heikinheimo:2018esa}. Second, in the region allowed by Bullet Cluster, the DM will be overproduced under the assumption of standard thermal history in the early Universe. However, entropy injection can dilute such overproduction bringing the relic density to an acceptable level. In End Matter  Fig.~\ref{fig:entropy_production}, we provide an estimate of the amount of entropy injection necessary to dilute the abundance to an acceptable level. It is therefore important to note that our analysis probes a wide variety of strongly-interacting DM scenarios without assuming any kinetic equilibrium provided some feeble interaction between the DS and SM exists. In fact, an absorption signal at IceCube-Gen2 Radio in this region would be a strong indication that the early Universe underwent non-standard thermal history. Third, the Bullet Cluster bound can be relaxed in some cases~\cite{Adhikari:2022sbh}, e.g. in presence of velocity-dependent interactions, anisotropic or dissipative scattering, baryonic feedback, or observational measurement bias, which could open up additional DM parameter space that can be probed using our absorption signal.   

\begin{figure}[t!]
    \centering
    \includegraphics[width=\linewidth]{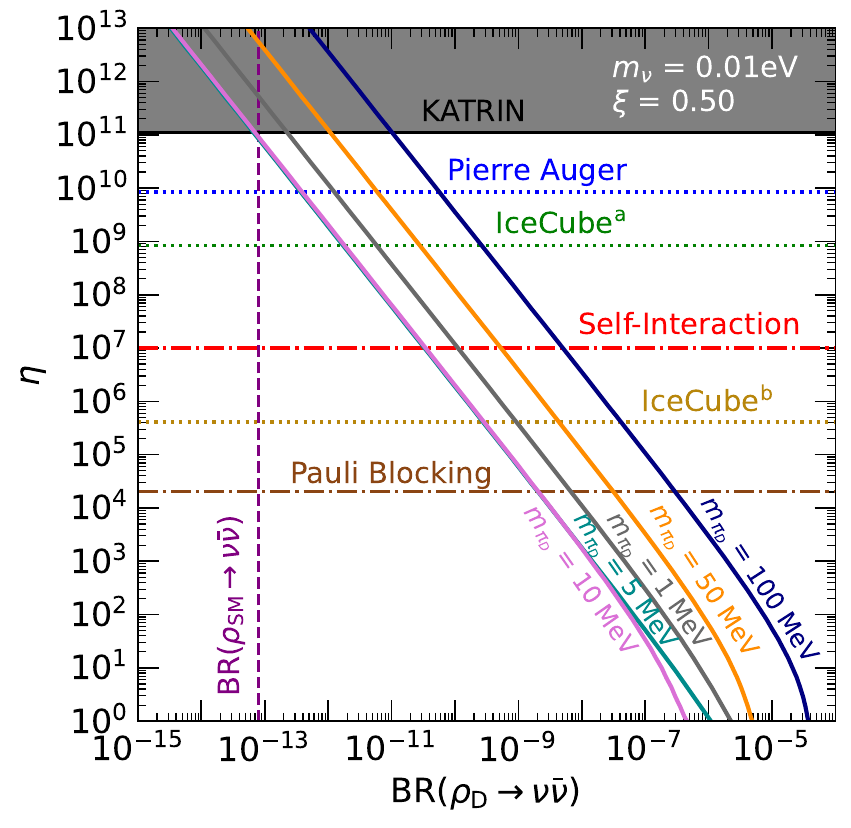}
    \caption{Minimum C$\nu$B overdensity $\eta$ that can be probed for a given BR into neutrinos, shown for different $m_{\pi_D}$ values. The per-flavor BR of the SM $\rho$ to neutrinos is shown for reference as the vertical dashed line. The gray shaded region is excluded by KATRIN~\cite{KATRIN:2022kkv}, while the other horizontal lines show other constraints from Pierre Auger~\cite{Zhang:2025rqh}, IceCube~\cite{Zhang:2025rqh, Herrera:2026pzj} observations, as well as theoretical limits from neutrino self-interaction~\cite{Smirnov:2022sfo} and Pauli blocking~\cite{Bondarenko:2023ukx} arguments. }
    \label{fig:Eta vs BR}
\end{figure}

We also examine, for specific choices of the DS model parameters $\xi$, $m_{\pi_D}$, ${\rm BR}(\rho_D\to \nu\bar\nu)$, what value of the neutrino overdensity $\eta$ would be required in order for an absorption signal to be detected by IceCube-Gen2. This is interesting because the C$\nu$B has so far evaded direct detection primarily due to the low neutrino energies, and its flux, local number density, and energy spectrum are not measured, but only loosely constrained by existing observations. In Fig.~\ref{fig:Eta vs BR}, we show 2$\sigma$ contours in the BR--$\eta$ plane for various $m_{\pi_D}$ values with a fixed $\xi=0.5$.  
We find maximum sensitivity for the smallest $\eta$ values around $m_{\pi_D}=5$--10 MeV. 
Irrespective of the DM mass, increasing the BR into neutrinos allows us to probe smaller overdensities up to the standard cosmological value of $\eta=1$. For comparison, we also show existing upper limits on $\eta$ from KATRIN~\cite{KATRIN:2022kkv} (gray-shaded), from the absence of cosmic-ray upscattered neutrinos at Pierre Auger~\cite{Zhang:2025rqh} and IceCube~\cite{Zhang:2025rqh, Herrera:2026pzj} (the two different IceCube limits are denoted by the superscripts `a' and `b', since there is no consensus yet), and Pauli blocking considerations~\cite{Bondarenko:2023ukx}, as well as the maximum overdensity that can be produced by a neutrino self-interaction model based on Yukawa coupling to a light scalar~\cite{Smirnov:2022sfo}. The per-flavor SM value of the BR [cf.~Eq.~\eqref{eq:BR} divided by 3] is shown as the vertical line for reference. It is clear from Fig.~\ref{fig:Eta vs BR} that neutrino-DS interactions can provide an  enhanced sensitivity to C$\nu$B.   

\begin{figure}
    \centering
    \includegraphics[width=\linewidth]{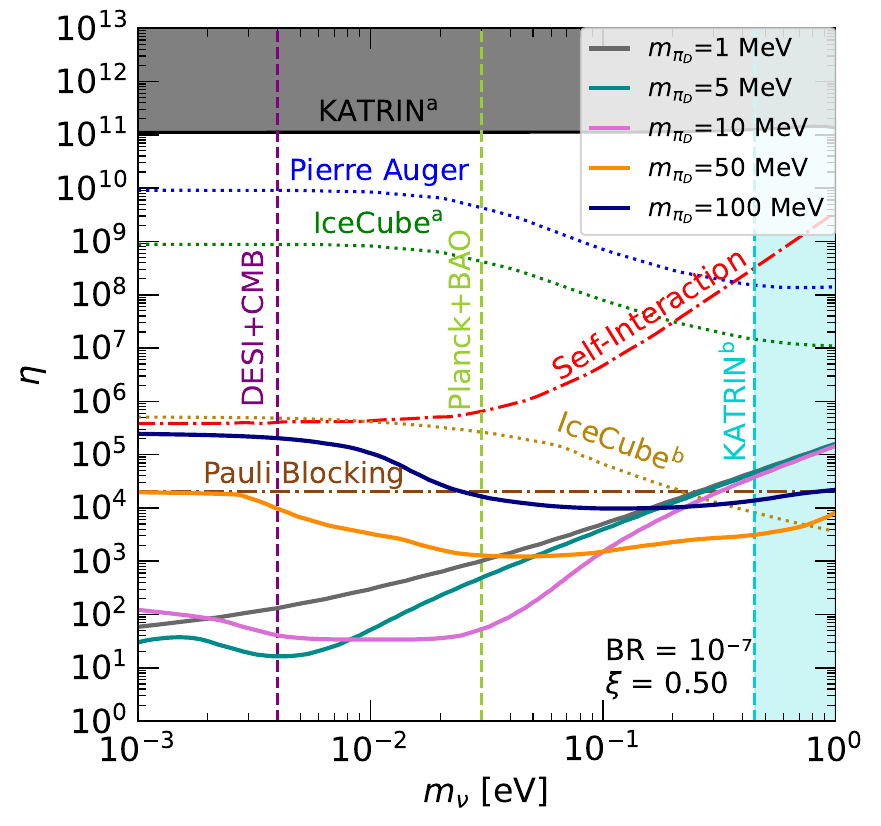}
    \caption{Minimum C$\nu$B overdensity that can be probed as a function of the lightest neutrino mass. The vertical lines show the laboratory and cosmology bounds on neutrino mass. The other labels are the same as in Fig.~\ref{fig:Eta vs BR}.}
    \label{fig:Eta vs m_nu}
\end{figure}

\begin{figure*}[t!]
    \centering
\includegraphics[width=0.49\linewidth]{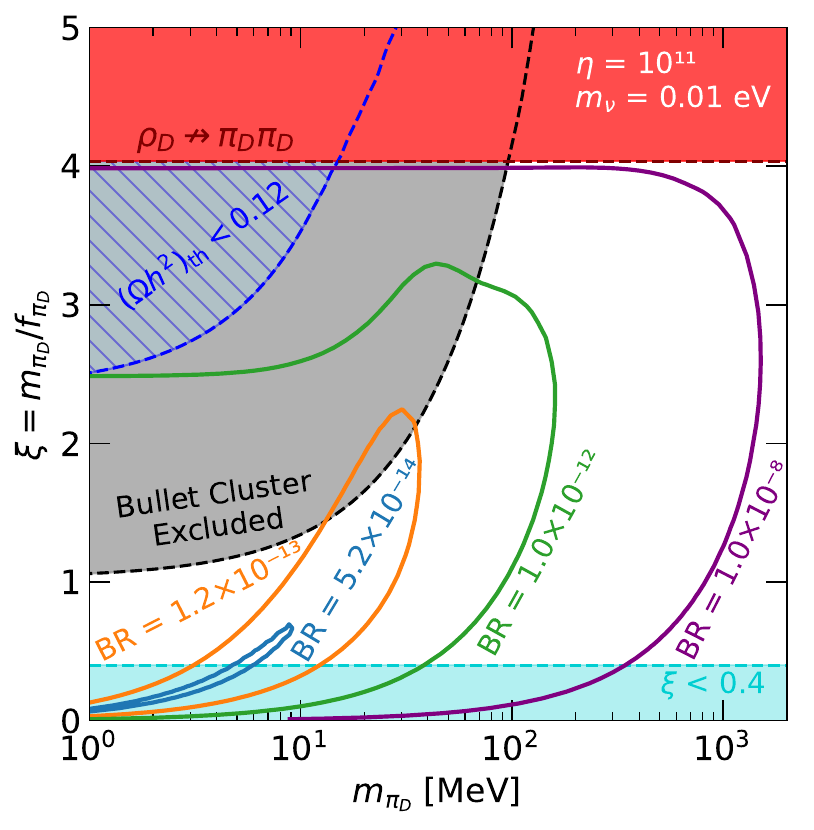}
\includegraphics[width=0.49\linewidth]{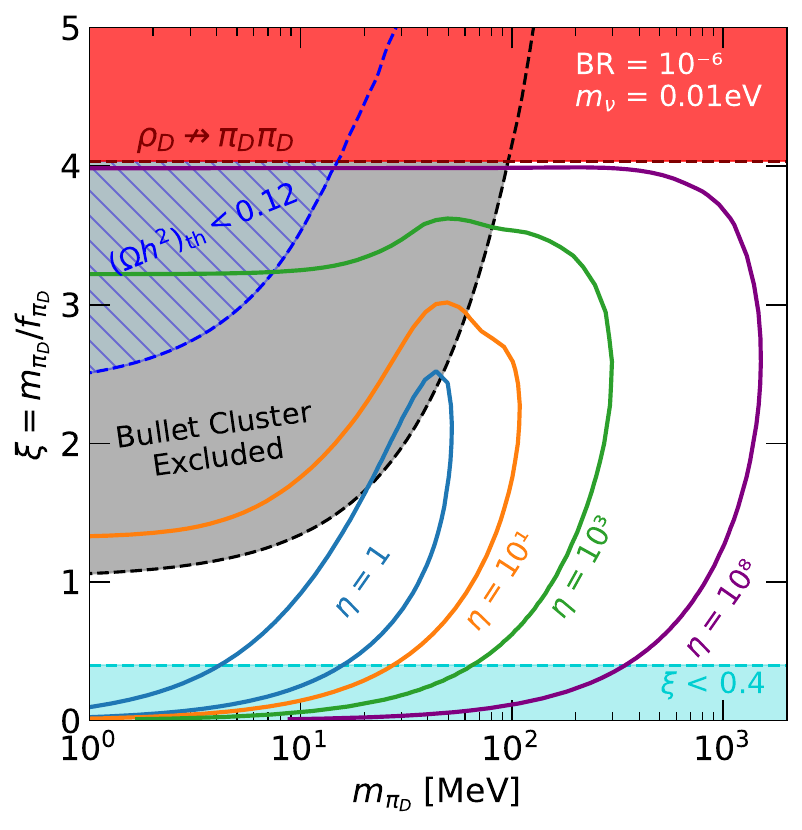}
    \caption{Similar to Fig.~\ref{fig:BR_contours-eta=1}, but with different values of $\eta$ and ${\rm BR}(\rho_D\to \nu\bar{\nu})$. {\bf Left panel:} $ \eta= 10^{11}$, the maximum value allowed by KATRIN~\cite{KATRIN:2022kkv}, with different BR contours. {\bf Right panel:} ${\rm BR}(\rho_D\to \nu\bar{\nu})=1.0\times10^{-6}$ with different $\eta$ contours. 
    }
\label{fig:BR_contours-eta=1e11 and Eta_contours}
\end{figure*}

We also explore in Fig.~\ref{fig:Eta vs m_nu} what neutrino mass gives the largest signature for the lowest overdensity across a range of $m_{\pi_D}$ with fixed BR and $\xi$. As in Fig.~\ref{fig:Eta vs BR}, the best sensitivity is obtained for $m_{\pi_D} = 5$--10 MeV for our specific choice of BR and $\xi$. This happens around $m_{\nu,{\rm lightest}} = 0.01$ eV. If the mass of the lightest neutrino is small enough, then the masses of the different mass eigenstates may be different enough that multiple distinct absorption features appear at different energies (cf.~Fig.~\ref{fig:spectrum}). As we increase the lightest neutrino mass, the fractional difference between the different masses is small, and the different absorption features merge into one, thus eroding the sensitivity. Once again, we include the upper limits on $\eta$ from KATRIN~\cite{KATRIN:2022kkv} (denoted as $\rm{KATRIN}^{a}$), Pierre Auger~\cite{Zhang:2025rqh} and IceCube~\cite{Zhang:2025rqh, Herrera:2026pzj}, Pauli blocking~\cite{Bondarenko:2023ukx} and neutrino self-interaction~\cite{Smirnov:2022sfo}. In addition, we show the KATRIN bound on the absolute neutrino mass~\cite{KATRIN:2024cdt} (denoted as $\rm{KATRIN}^{b}$), as well as cosmological bounds from Planck~\cite{Planck:2018vyg} and recent DESI+CMB~\cite{DESI:2025zgx} data, which depend on the cosmological datasets used (and therefore, not shaded). Again, it is clear that we can probe C$\nu$B overdensities well beyond existing constraints using the neutrino-DS interactions.

Although we have used IceCube-Gen2 Radio array~\cite{IceCube-Gen2:2020qha} to show the sensitivity projections, our analysis can be extended to other current/future ultra-high-energy neutrino experiments, such as RNO-G~\cite{RNO-G:2020rmc}, GRAND~\cite{GRAND:2018iaj}, BEACON~\cite{Wissel:2020sec}, HERON~\cite{GRAND:2025rps}, TAMBO~\cite{Arguelles:2026btb}, POEMMA~\cite{POEMMA:2020ykm} and PUEO~\cite{PUEO:2020bnn} as well. 

\acknowledgments
\section*{Acknowledgments}
We thank Brian Clark for useful discussion  on IceCube-Gen2 Radio sensitivity. C.V.C. was generously supported by Washington University in St.~Louis through the Edwin Thompson Jaynes
Postdoctoral Fellowship. B.D. was supported in part by the U.S. Department of Energy under grant No.
DE-SC0017987. S.~K. is supported by the FWF project number P 36947-N. B.D. and S.K. thank the organizers of PPC 2024 at IIT, Hyderabad where this work was initiated. We acknowledge the use of OpenAI's {\tt GPT-5.6 Luna} and Anthropic's {\tt Sonnet 4.6} to increase the efficiency of our numerical code.

\section{End Matter}\label{sec:Appendix}

\begin{figure}[t!]
    \centering
    \includegraphics[width=\linewidth]{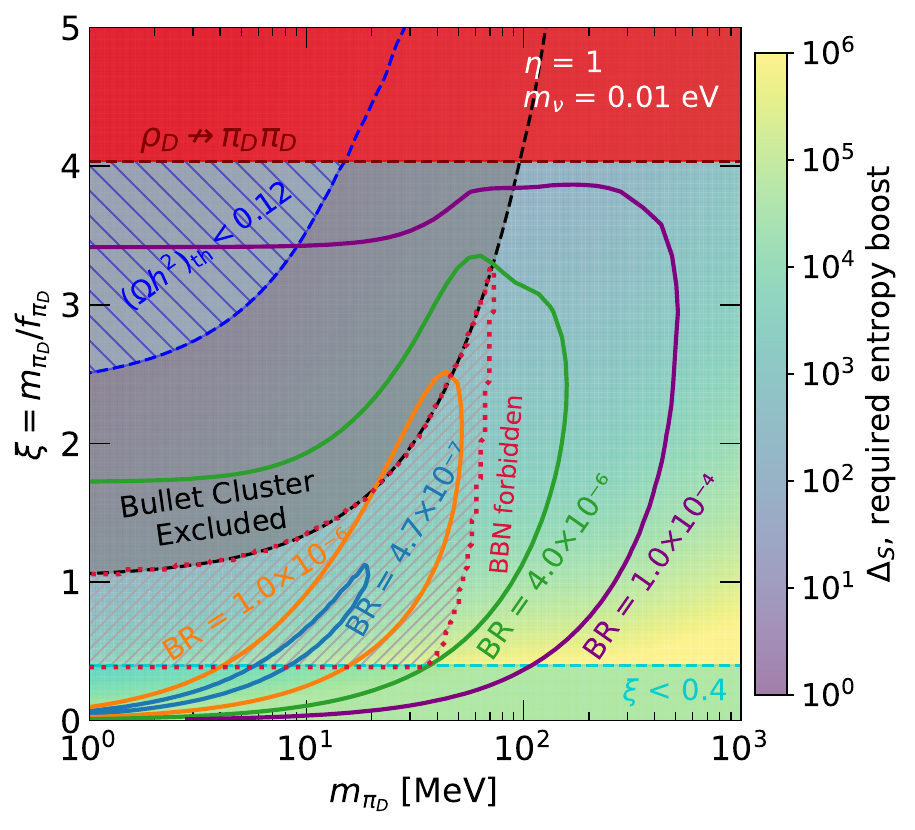}
    \caption{Amount of entropy production required to dilute DM overabundance in the viable region. We also illustrate region where DM freezes out after BBN has already started ($T_f < T_{\rm BBN})$. We take $T_f = 4\,\rm{MeV}$. The other contours are the same as in Fig.~\ref{fig:BR_contours-eta=1}.}
    \label{fig:entropy_production}
\end{figure}

The results presented in the main text assume standard C$\nu$B number density and no overdensity, i.e.~$\eta = 1$. However, since the direct laboratory constraint on local overdensity is rather weak:  $\eta\lesssim 10^{11}$~\cite{KATRIN:2022kkv} and the astrophysical constraints have large uncertainties~\cite{DeMarchi:2024zer, Zhang:2025rqh, Herrera:2026pzj, Azeredo:2026qnc}, in this section we redo the analysis with $\eta>1$ to see how much improvement can be achieved in the sensitivity. In Fig.~\ref{fig:BR_contours-eta=1e11 and Eta_contours}  left panel, we show the results for $\eta = 10^{11}$, representing the most optimistic scenario. In this case, we see that IceCube-Gen2 Radio will have sensitivity to BRs even below the SM expectation. The right panel shows the same parameter space, but with the BR held at $1.0\times10^{-6}$ and contours denoting different values of $\eta$. We see that a significant portion of parameter space can be probed even for $\eta = 1$, and this region becomes further enlarged in the presence of an overdensity.

In Fig.~\ref{fig:entropy_production}, we reexamine the parameter space shown in Fig.~\ref{fig:BR_contours-eta=1} and show the amount of entropy production required to dilute the DM relic density in the viable region. Entropy production in the early Universe has been well studied in a number of contexts; for recent studies, see e.g., Refs.~\cite{Evans:2019jcs, Asadi:2021bxp}. As our DM candidate is light, and it freezes out when non-relativistic, it is possible that DM freezes out around the time of Big-Bang Nucleosynthesis (BBN), in which case entropy production must happen as BBN is in progress. Such low reheating scenarios are  ruled out given the BBN constraints. Following Ref.~\cite{Hannestad:2004px}, we take the upper bound on reheating temperature to be $4$ MeV. Thus, if DM freeze out at temperature $T_f < 4$ MeV, we consider that to be a forbidden region, which is labeled as ``BBN forbidden" in the plot. This still leaves a large region in the  $m_{\pi_D}$--$\xi$ plane where DM may be overproduced. We show that the entropy must be diluted by a factor of $\Delta_S=10^2$--$10^6$ for such overproduced DM to become compatible with the observed relic density.

\bibliography{main}

\end{document}